\documentclass[preprintnumbers, floatfix, twocolumn,
preprintnumbers, PRD, superscriptaddress,nofootinbib]{revtex4}
\usepackage{eurosym}
\usepackage{amsfonts}
\usepackage{amsmath}
\usepackage{amssymb,epsf}
\usepackage{latexsym}
\usepackage{graphicx,epsfig}
\usepackage{amssymb}
\usepackage{subfigure}
\usepackage{epstopdf}
\usepackage[colorlinks=true,citecolor=blue,linkcolor=blue,urlcolor=black]{hyperref}
\usepackage{color}

\graphicspath{{Images/}}

\renewcommand{\&}{\textup{\symbol{`\&}}}

\begin{document}

\title{Microstructure of charged AdS black hole via $P-V$ criticality}
\author{Amin Dehyadegari}
\affiliation{Physics Department and Biruni Observatory, Shiraz University, Shiraz 71454,
Iran}
\author{Ahmad Sheykhi}
\email{asheykhi@shirazu.ac.ir}
\affiliation{Physics Department and Biruni Observatory, Shiraz University, Shiraz 71454,
Iran}
\affiliation{Max-Planck-Institute for Gravitational Physics (Albert-Einstein-Institute),
14476 Potsdam, Germany}
\author{Shao-Wen Wei}
\email{ weishw@lzu.edu.cn}
\affiliation{Institute of Theoretical Physics \& Research Center of Gravitation, Lanzhou
University, Lanzhou 730000, People's Republic of China}

\begin{abstract}
We suggest a new thermodynamic curvature, constructed via
adiabatic compressibility, for examining the internal
microstructure of charged black holes in an anti-de Sitter (AdS)
background. We analyze the microscopic properties of small-large
phase transition of black holes with pressure and volume as the
fluctuation variables. We observe that strong repulsive
interactions dominate among the micro-structures of near extremal
small black holes, and the thermodynamic curvature diverges to
positive infinity for the extremal black holes. At the critical
point, however, thermodynamic curvature diverges to negative
infinity.
\end{abstract}

\maketitle


\section{Introduction \label{Intro}}

\bigskip Phase transition is a fascinating phenomenon in black holes
thermodynamics which has received considerable attentions in recent years.
This is mainly motivated by AdS/CFT duality, which states that there exists
a correspondence black holes in asymptotically anti-de Sitter (AdS)
spacetime and the conformal field theory living on its boundary \cite%
{HP,Witten}. A significant interest has been arisen for study phase
transition of AdS black hole in an extended phase space in which the
cosmological constant can be regarded as thermodynamic pressure which can
vary \cite{Dolan,PV}. In this viewpoint, the mass of black hole is
identified as the enthalpy \cite{enthalpy}. It was shown \cite{PV} that the
four dimensional charged AdS black hole demonstrates the first order
(discontinuous) and second order (continuous) phase transitions between the
small and large black holes in an extended phase space. This phase
transition is analogous to the Van der Waals gas/liquid phase transition,
thus, their critical exponents are the same as well. The investigation on
the critical behavior of black holes in this context is often referred to as
\textquotedblleft $P$\textit{-}$V$\textit{\ criticality}" and has widely
explored in the literatures \cite%
{Hendi,Sherkat,Sherkat1,Rabin,Kamrani,DSD1,DSD2,DSDH,Dehghani,BIRPT} and
references therein. Some interesting phenomena have been observed in the
extended phase space of black holes such as \textit{zeroth order} phase
transition \cite{NAAA} and reentrant phase transition \cite{KerrRPT} as well
as triple critical point \cite{triplep} as well as superfluid like phase
transition \cite{superfluidBH}. More recently, a universality class of the
critical behavior of AdS black holes in an extended phase space has been
studied by a general approach without specifying the functional form of the
spacetime metric \cite{Majhi}.

An alternative approach to investigate critical behavior of black holes is
to consider the electric charge ($Q$) of the black hole as a thermodynamical
variable while keeping the cosmological constant as a fixed parameter. From
the physical point of view, the electric charge of black hole is a natural
variable which can take on arbitrary values and it affects the thermodynamic
properties of AdS black hole. In this case, it was argued \cite{VDW1} that
there exists a small-large black hole phase transition for the charged black
hole in a fixed AdS background. It has been demonstrated \cite{AAA} that
this phase transition is physically conventional in an alternative phase
space where the square of the electric charge ($Q^{2}$) is viewed as an
independent thermodynamic variable of the black hole system. In this
perspective, the new thermodynamic response function correctly signifies
stable and unstable regimes and the critical behavior of the black hole
resembles with Van der Waals fluid, belonging to the same universality class
\cite{AMajhi}. Phase transition of black holes in an alternative phase space
have been explored in different setups \cite{DS,Homa,Arab}. More recently,
the authors of Ref. \cite{ASheykhi} investigated thermodynamic phase
structure of Born-Infeld and charged dilaton \cite{ASheykhi2} black holes in
a fixed AdS spacetime by studying the behavior of specific heat.

The theory of covariant thermodynamic fluctuations provides a powerful
geometric framework to study properties of underlying thermal system,
completely from the thermodynamic viewpoint \cite{Rup0,Rup1}. In this
context, Ruppeiner defined the Riemannian metric on the equilibrium
thermodynamic state space as the second derivatives of entropy. In his
series of works \cite{Rup2,Rup3,Rup4}, it has been confirmed that
thermodynamic curvature (Ricci scalar) arising out of a such metric is
related to the microscopic interactions, where the thermodynamic curvature
is positive (negative) for the repulsive (attractive) interaction. In
addition, thermodynamic curvature diverges at the critical point for pure
fluid systems. With regard to this approach, the microscopic behavior and
phase transition of various kinds of black holes have been explored \cite%
{Rup5,Rup6,Rup7,comment}. In all these works, thermodynamic curvature has a
finite value at the critical point. Recently, a new normalized thermodynamic
curvature was proposed to understand the microscopic behavior of charged AdS
black hole in an extended phase space where the temperature and volume are
treated as fluctuating variables \cite{Wei1,Wei2,Wei3}. In this formalism,
thermodynamic curvature is normalized with respect to the heat capacity at
constant volume. It was shown that the microstructure of small black hole
has a weak repulsive interaction and the thermodynamic curvature goes to
infinity at the critical point of phase transition.

In this paper, we offer a new thermodynamic curvature, which is constructed
via the adiabatic compressibility, for examining the internal microstructure
of charged AdS black holes in an extended phase space with fixed charge. In
particular, we analyze the microscopic properties of small-large phase
transition of black holes with pressure and volume as the fluctuation
variables. Our work differs from \cite{Wei1,Wei2} in that we allow the
pressure and volume to fluctuate and normalize the thermodynamic curvature
by the adiabatic compressibility, while the authors of \cite{Wei1,Wei2}
considered the temperature and volume as the fluctuating quantities and
normalized the thermodynamic curvature by the heat capacity at constant
volume. We observe that strong repulsive interactions dominate among the
micro-structures of small black holes where the thermodynamic curvature
diverges to positive infinity. It is shown that the thermodynamic curvature
diverges to negative infinity at the critical point.

The structure of the paper is laid out as follows. We begin in Sec. \ref%
{Review} by giving a brief review of the thermodynamics and critical
behavior of the four dimensional charged AdS black holes in the extended
phase space. In Sec. \ref{Rup}, we first introduce the Ruppeiner geometry
and obtain the corresponding line element for a thermodynamic system in
terms of the entropy and pressure. Then, we use the thermodynamic curvature
to investigate in detail the microstructure of charged AdS black hole.
Finally, we present some remarks in Sec. \ref{FR}.

\section{Thermodynamics and phase transition of charged AdS black holes\label%
{Review}}

We start with a brief review on the thermodynamics properties and $P-V$
criticality of Reissner-Nordstrom (RN)-AdS black hole in an extended phase
space. The action of Einstein-Maxwell theory in four-dimensional spacetime
with a cosmological constant ($\Lambda $) is
\begin{equation}
I=\frac{1}{16\pi }\int d^{4}x\sqrt{-g}\left( \mathcal{R}\text{ }-2\Lambda
-F_{\mu \nu }F^{\mu \nu }\right) ,  \label{Act1}
\end{equation}%
where $\mathcal{R}$ is the scalar Riemann curvature, $F_{\mu \nu }$ is the
electromagnetic field strength that is defined as $F_{\mu \nu }=\partial
_{\mu }A_{\nu }-\partial _{\nu }A_{\mu }$ with the gauge field $A_{\mu }$.
The negative cosmological constant $\Lambda $ is related to the AdS radius $%
L $ by the relation, $\Lambda =-3/L^{2}$. In four dimensions, the line
element of the spherically symmetric RN-AdS metric is given by \cite{PV}
\begin{eqnarray}
ds^{2} &=&-f(r)dt^{2}+\frac{dr^{2}}{f(r)}+r^{2}d\Omega ^{2}, \\
f(r) &=&1-\frac{2M}{r}+\frac{Q^{2}}{r^{2}}+\frac{r^{2}}{L^{2}},
\end{eqnarray}%
where $d\Omega ^{2}$ is the metric of the unit two sphere. Herein, the
parameters $M$ and $Q$ are, respectively, the mass and charge of black hole
where the position of the black hole event horizon ($r_{+}$) is determined
as a largest positive real root of $f(r_{+})=0$. The only nonvanishing
component of the electromagnetic field tensor is given by $F_{tr}=Q/r^{2}$.

The Hawking temperature of the RN-AdS black hole on an event horizon is
obtained as \cite{PV}
\begin{equation}
T=\frac{f^{\prime }(r_{+})}{4\pi }=\frac{1}{4\pi r_{+}}\left( 1+\frac{%
3r_{+}^{2}}{L^{2}}-\frac{Q^{2}}{r_{+}^{2}}\right) ,  \label{temp}
\end{equation}%
and the entropy is
\begin{equation}
S=\pi r_{+}^{2}.  \label{entropy}
\end{equation}%
By interpreting the cosmological constant as a thermodynamic pressure, $%
P=-\Lambda /(8\pi )$, and its conjugate quantity as a black hole
thermodynamic volume, $V=4\pi r_{+}^{3}/3$, the first law of black hole
thermodynamics and the corresponding Smarr formula take the form,
respectively,
\begin{eqnarray}
dM &=&TdS+VdP+\Phi dQ,  \label{Msm} \\
M &=&2TS+\Phi Q-2VP,
\end{eqnarray}%
where $\Phi =Q/r_{+}$ is the electric potential measured with respect to the
event horizon. In this consideration, the mass ($M$) of the black hole is
identified as the enthalpy. Also, the thermodynamic process is carried out
in the extended phase space. It is worthwhile to mention that according to
Eq. (\ref{entropy}) and black hole thermodynamic volume formula, the entropy
is only a function of area/volume, i.e. $S=S\left( V\right) $. This feature
of the black hole will be used in the next section.
\begin{figure}[t]
\epsfxsize=8.5cm \centerline{\epsffile{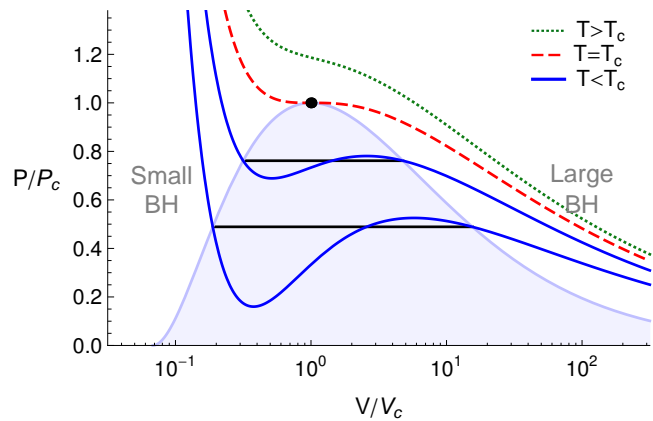}}
\caption{{}$P$-$V$ diagram of RN-AdS black holes. The region of the first
order phase transition is identified where the isobars (black horizontal
lines) remedy the unstable regime by the Maxwell equal area law. The areas
above and below the black isobar are equal one another which is not seen
because of logarithmic scale on the horizontal axis. The critical point is
marked by a black spot. Note the logarithmic scale on the horizontal axis.}
\label{fig1}
\end{figure}

For the four-dimensional charged AdS black hole, the equation of state, $%
P=P\left( T,V\right) $, is obtained by using Eq.(\ref{temp}) as
\begin{equation}
P=\frac{T}{2r_{+}}-\frac{1}{8\pi r_{+}^{2}}+\frac{Q^{2}}{8\pi r_{+}^{4}},
\label{EoS}
\end{equation}%
where $r_{+}=\left( 3V/4\pi \right) ^{1/3}$. The behavior of isotherms in
the $P$-$V$ diagram is shown in Fig. \ref{fig1}. We see that the critical
point is an inflection point on the isotherm which is characterized by
\begin{equation}
\frac{\partial P}{\partial V}\Big|_{T_{c}}=0,\quad \quad \quad \frac{%
\partial ^{2}P}{\partial V^{2}}\Big|_{T_{c}}=0.  \label{CP}
\end{equation}%
One obtains the critical quantities as
\begin{equation}
T_{c}=\frac{\sqrt{6}}{18\pi Q},\quad P_{c}=\frac{1}{96\pi Q^{2}},\quad
V_{c}=8\sqrt{6}\pi Q^{3}.  \notag
\end{equation}%
For $T<T_{c}$, an oscillating part of the isotherm denotes unstable region
where the isothermal compressibility is negative, i.e.
\begin{equation}
\kappa _{{}_{T}}=-\frac{1}{V}\left. \frac{\partial V}{\partial P}\right\vert
_{T}<0.
\end{equation}%
This instability is replaced by an isobar (the horizontal line) via the
Maxwell equal area construction, $\oint VdP=0$, which means that there
exists a first order phase transition between the small black hole and large
black hole. The small-large black hole transition region, determined by
Maxwell construction, has the following forms \cite{Smailagic}%
\begin{eqnarray}
\widetilde{T}^{2} &=&\widetilde{P}(3-\sqrt{\widetilde{P}})/2,  \notag \\
\widetilde{P} &=&\frac{7+6\widetilde{V}^{2/3}-4\sqrt{3+6\widetilde{V}^{2/3}}%
}{\widetilde{V}^{4/3}},
\end{eqnarray}%
where the reduced thermodynamic variables are defined as
\begin{equation}
\widetilde{T}=\frac{T}{T_{c}},\quad \widetilde{P}=\frac{P}{P_{c}},\quad
\widetilde{V}=\frac{V}{V_{c}}.  \notag
\end{equation}%
It is worthwhile to note that the two phase of small and large black holes
cannot be distinguished above the critical point. In the next section, we
examine the behavior of charged black hole in the Ruppeiner geometry.


\section{Ruppeiner geometry \label{Rup}}

In this section, we apply the concept of the Ruppeiner thermodynamic
geometry as a useful tool to study the microscopic structure of charged AdS
black holes. The Ruppeiner geometry arises from the Gaussian thermodynamic
fluctuation theory which is constructed on the thermodynamic state space
\cite{Rup1}. In two dimensions, the Riemannian curvature scalar, $R$,
(thermodynamic curvature) gives complete information about the Ruppeiner
geometry which is connected with the inter-particle interaction in a
thermodynamic system. Specially, the positive (negative) sign of the
thermodynamic curvature indicates the repulsive (attractive) interaction,
while $R=0$ corresponds to no interaction \cite{Rup2,Rup3,Rup4}. In the
following, we first derive the thermodynamic fluctuation metric in the ($S$,$%
P$) coordinates, where a thermodynamic potential is the enthalpy. Then,
using the fact that the entropy of the charged AdS black hole only depends
on the volume, we investigate the thermodynamic curvature of black hole
through the ($P$,$V$) plane.

\subsection{Ruppeiner metric \label{RupM}}

Consider a thermodynamic system characterized by the entropy ($S$), internal
energy ($U$) and volume ($V$) such that the line element between two
thermodynamic states is \cite{Rup1}
\begin{equation}
\Delta l^{2}=g_{\mu \nu }\Delta x^{\mu }\Delta x^{\nu },  \label{metric1}
\end{equation}%
where $x^{\mu }=\left( U,V\right) $ and the metric element $g_{\mu \nu }$ is
given by
\begin{equation}
g_{\mu \nu }=-\frac{\partial ^{2}S}{\partial x^{\mu }\partial x^{\nu }}.
\notag
\end{equation}%
In the entropy representation, the first law of thermodynamics for this
system is expressed as follows
\begin{equation}
dS=\frac{1}{T}dU+\frac{P}{T}dV,  \label{firstlaw}
\end{equation}%
where $T$ and $P$ are temperature and pressure, respectively. Using the
first law of thermodynamics Eq.(\ref{firstlaw}), the line element Eq.(\ref%
{metric1}) can be written as
\begin{equation}
\Delta l^{2}=\frac{1}{T}\Delta T\Delta S-\frac{1}{T}\Delta P\Delta V.
\label{metric2}
\end{equation}%
To express the above line element in ($S$,$P$) coordinates, we have
\begin{eqnarray}
\Delta T &=&\frac{\partial T}{\partial S}\Big|_{P}\Delta S+\frac{\partial T}{%
\partial P}\Big|_{S}\Delta P,  \notag \\
\Delta V &=&\frac{\partial V}{\partial S}\Big|_{P}\Delta S+\frac{\partial V}{%
\partial P}\Big|_{S}\Delta P.  \label{metric3}
\end{eqnarray}%
Substituting Eqs.(\ref{metric3}) into Eq.(\ref{metric2}) and using the
Maxwell relation
\begin{equation}
\frac{\partial T}{\partial P}\Big|_{S}=\frac{\partial V}{\partial S}\Big|%
_{P},  \notag
\end{equation}%
one obtains the thermodynamic line element
\begin{equation}
\Delta l^{2}=\frac{1}{C_{P}}\Delta S^{2}+\frac{V}{T}\kappa _{S}\Delta P^{2},
\label{metric4}
\end{equation}%
where $C_{P}=T\left( \partial S/\partial T\right) _{P}$ is the heat capacity
at constant pressure and $\kappa _{S}=-1/V\left( \partial V/\partial
P\right) _{S}$ is the adiabatic compressibility. Here, the thermodynamic
potential is the enthalpy where the independent variables are entropy and
pressure.
\subsection{Thermodynamic curvature in $P$-$V$ diagram \label{Thermodynamic
Rupp}}

Now, we use Eq. (\ref{metric4}) to investigate microstructure of RN-AdS
black hole in an extended phase space. Due to the fact that the entropy of
black hole only depends on the volume, i.e. $S=(3^{2}\pi
/2^{4})^{1/3}V^{2/3} $, the line element of the Ruppeiner geometry can be
written as
\begin{equation}
\Delta l^{2}=\frac{1}{C_{P}}(\frac{\pi }{6V})^{2/3}\Delta V^{2}+\frac{V}{T}%
\kappa _{S}\Delta P^{2},  \label{metric5}
\end{equation}%
where the pressure and volume are taken as the fluctuation variables. For
the black hole, the adiabatic compressibility ($\kappa _{S}$) vanishes
similar to the heat capacity at constant volume, i.e. $C_{V}=T\left(
\partial S/\partial T\right) _{V}=0$ \footnote{%
The entropy of the Van der Waals fluid system is a function of the
temperature and volume i.e. $S=S\left( T,V\right) $ \cite{Wei2,Landau}. This
would imply that the adiabatic compressibility is non-zero ($\kappa _{S}\neq
0)$ and it has a finite value at the critical point.}. Hence, following \cite%
{Wei1,Wei2}, we define a normalized thermodynamic curvature, $R_{N}$, based
on the adiabatic compressibility
\begin{equation} \label{NTC}
R_{N}=R\kappa _{S}.
\end{equation}%
In what follows, we analyze in detail the behavior of the normalized
thermodynamic curvature as function of the pressure and volume. By
performing simple calculations, we obtain the normalized thermodynamic
curvature
\begin{equation}
R_{N}=\frac{16\widetilde{V}^{2/3}(3\widetilde{V}^{2/3}-1)(5-6\widetilde{V}%
^{2/3}+9\widetilde{P}\widetilde{V}^{4/3})}{(1-2\widetilde{V}^{2/3}+%
\widetilde{P}\widetilde{V}^{4/3})^{2}(1-6\widetilde{V}^{2/3}-3\widetilde{P}%
\widetilde{V}^{4/3})},  \label{RN}
\end{equation}%
\begin{figure}[t]
\epsfxsize=8.5cm \centerline{\epsffile{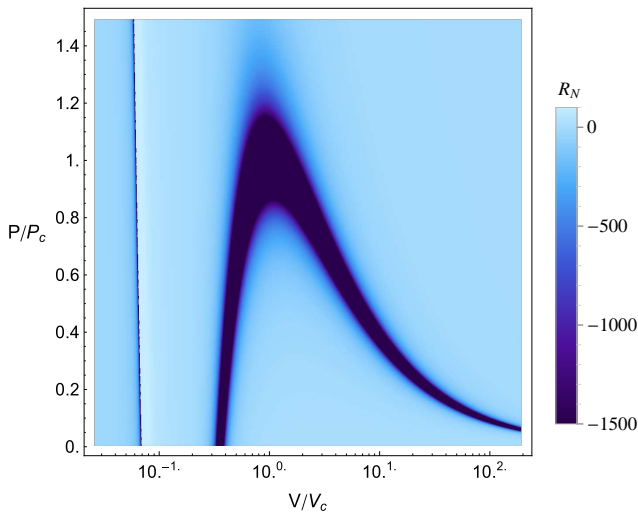}}
\caption{{}The normalized thermodynamic curvature as a function of the
pressure $P$ and volume $V$. Note the logarithmic scale on the horizontal
axis.}
\label{fig2}
\end{figure}

\begin{figure}[t]
\epsfxsize=8.5cm \centerline{\epsffile{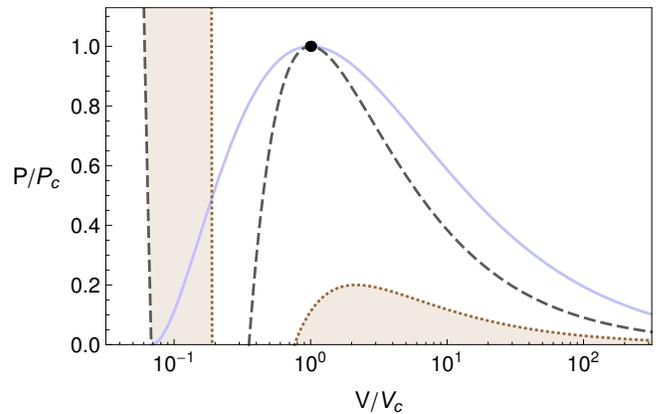}}
\caption{{}Transition curve (light blue solid line), vanishing curve (brown
dotted line) and diverging curve (gray dashed line) of $R_{N}$. The region
identified corresponds to positive $R_{N}$, otherwise $R_{N}$ is negative.
Both the transition and diverging curves start at $\widetilde{V}=1/6\protect%
\sqrt{6}$. The region to the left of the shaded region on the left
side of the figure is excluded becuase temperature is negative. The
critical point is marked by a black spot. Note the logarithmic scale on the
horizontal axis.}
\label{fig3}
\end{figure}
which is expressed in terms of the reduced thermodynamic variables.
Remarkably, the $R_{N}$ is independent of the charge of a black hole in Eq.(%
\ref{RN}). It should be noted that if one uses Eq.(\ref{metric4}) instead of
Eq.(\ref{metric5}) for the Ruppeiner line element, the normalized
thermodynamic curvature ($R_{N}$), Eq.(\ref{RN}), does not change. The
overall behavior of the normalized thermodynamic curvature as a function of $%
P/P_{c}$ and $V/V_{c}$ is illustrated in Fig.\ref{fig2}. As can be
ascertained from Fig.\ref{fig2}, the $R_{N}$ goes to negative infinity in
certain regions of the plane. From Eq.(\ref{RN}), $R_{N}$ diverges along the
curves
\begin{eqnarray}
\widetilde{P}_{div} &=&\frac{2\widetilde{V}^{2/3}-1}{\widetilde{V}^{4/3}},
\label{div1} \\
\widetilde{P}_{div} &=&\frac{1-6\widetilde{V}^{2/3}}{3\widetilde{V}^{4/3}},
\label{div2}
\end{eqnarray}%
The divergent curve in Eq. (\ref{div2}) corresponds to the
extremal black holes which are at zero temperature. On the other
hand, $R_{N}$ obviously vanishes at the following curves
\begin{eqnarray}
\widetilde{P}_{0} &=&\frac{6\widetilde{V}^{2/3}-5}{9\widetilde{V}^{4/3}},
\notag \\
\widetilde{V}_{0} &=&\frac{1}{3\sqrt{3}},  \label{vanishing}
\end{eqnarray}%
where the dominant interaction between the microstructure of charged black
hole changes from attractive to repulsive and vice versa.

To better understand the behavior of the normalized thermodynamic curvature,
we show the diverging (gray dashed line) and vanishing (brown dotted line)
curves corresponding to Eqs.(\ref{div1}), (\ref{div2}) and (\ref{vanishing}%
), respectively, as well as the small-large black hole phase transition
(light blue solid line) curve in Fig.\ref{fig3}. In Fig.\ref{fig3}, the
critical point is highlighted by a black solid circle and the shaded regions
have positive values for $R_{N}$ which imply the domination of repulsive
interaction. In the other region, $R_{N}$ is negative which means the
microstructure interactions are attractive. As evident from Fig.\ref{fig3}, $%
R_{N}$ is negative for the large black hole, while there is a certain range
of volume in the small black hole phase ($\widetilde{V}<1$) that has
positive $R_{N}$. In this positive region, $R_{N}$ diverges to positive
infinity when the gray dashed line is approached from large values of
volume. i.e, the microstructure interaction of the small black hole is
strongly repulsive. A strongly repulsive interaction also exists in the
higher pressure regime (above the critical point) at low volume $\widetilde{V%
}$. The white region to the left of the gray dashed curve on the
left side of the Fig. \ref{fig3}, where black holes are
sufficiently small, is excluded due to the fact that temperature
is negative. Since the equation of state (\ref{EoS}) may not hold
in the transition region (below the light blue solid curve),
$R_{N}$ does not give any information about the black hole
microstructure. Furthermore, as also seen in Fig.\ref{fig3}, light
blue solid and gray dashed curves coincide at the critical point
where the thermodynamic functions of charged black hole are
characterized by a set of critical exponents \cite{PV}. Hence, the
normalized thermodynamic curvature diverges to negative infinity
($R_{N}\rightarrow -\infty $) at the critical point. This
situation is analogous to fluid in the critical point regime, such
as Van der Waals system \cite{Rup1,Wei1,Wei2}, where thermodynamic
curvature goes to negative infinity at the critical point.

To obtain an explicit expression of $R_{N}$ near the critical point, we
expand $R_{N}$, Eq.(\ref{RN}), around the critical point using Eq.(\ref{EoS}%
)
\begin{equation}
R_{N}=-\frac{9}{2}t^{-2},
\end{equation}%
\begin{figure}[t]
\epsfxsize=8.5cm \centerline{\epsffile{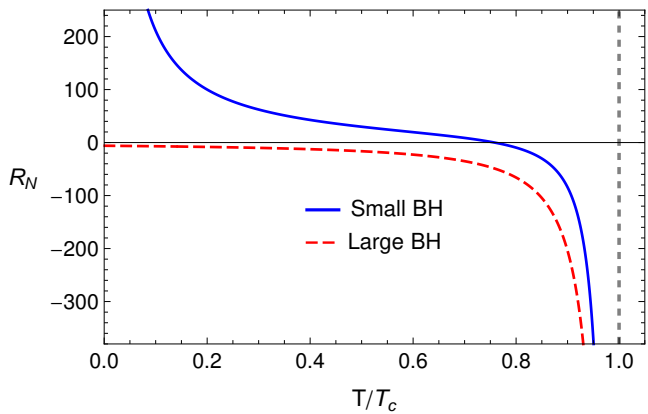}}
\caption{{} The normalized thermodynamic curvature $R_{N}$ for charged AdS
black hole along the transition curve in the small and large black holes
phases. $R_{N}$ of the small black hole changes the sign to positive at $%
\widetilde{T}=3\protect\sqrt{3}(7-3\protect\sqrt{5})/2\approx 0.7581$.}
\label{fig4}
\end{figure}
where $t=1-\widetilde{T}$ is the deviation from the critical temperature.
Therefore, $R_{N}$ has the universal critical exponent $2$ and critical
amplitude $-9/2$. Further, it is interesting to investigate the behavior of $%
R_{N}$ on the transition curve. In this respect, we plotted in Fig.\ref{fig4}%
, $R_{N}$ along the transition curve in both the small and the large black
holes phases from the critical temperature to zero. One observes from Fig.%
\ref{fig4} that $R_{N}$ in both phases diverges to $-\infty $ at the
critical temperature. In the large black hole phase, $R_{N}$ uniformly
negative and $\left\vert R_{N}\right\vert $ decreases as the temperature
decreases from the critical temperature, which it is small at $\widetilde{T}%
=0$. While, in the small black hole phase, $R_{N}$ changes sign and becomes
positive below $\widetilde{T}=3\sqrt{3}(7-3\sqrt{5})/2\approx 0.7581$.
Remarkably, $R_{N}$ diverges to positive infinity as $\widetilde{T}$ tends
to zero where strong repulsive interactions dominate.

\section{Final Remarks \label{FR}}

In this paper, we proposed a new thermodynamic curvature, by using
the adiabatic compressibility, for examining the internal
microstructure of charged AdS black holes in an extended phase
space. We explored the microscopic properties of small-large black
holes phase transition by considering the pressure and volume as
the fluctuation variables. We defined a normalized thermodynamic
curvature, $R_{N}=\kappa_{S} R$, where $\kappa_{S}$ is the
adiabatic compressibility, and studied the behavior of $R_{N}$  as
a function of the pressure and volume. The sign of $R_{N}$
determines the repulsive or attractive feature of black holes
microstructure. When $R_{N}>0$, the repulsive interaction
dominates, while $R_{N}<0$ indicates that the microstructure
interactions are attractive. We also observed that a strongly
repulsive interaction exists in the higher pressure regime (above
the critical point) at low volume. At the critical point however,
we have $R_{N}\rightarrow -\infty $, which is analogous to the Van
der Waals fluid in its critical point regime.


\begin{acknowledgments}
We are grateful to the Research Council of Shiraz University. A.S.
thanks Hermann Nicolai and Max-Planck-Institute for Gravitational
Physics, for hospitality. This work was supported by the National
Natural Science Foundation of China (Grant No. 11675064) and the
Fundamental Research Funds for the Central Universities (Grant No.
lzujbky-2019-it21).
\end{acknowledgments}


\end{document}